\documentclass[12pt]{article}
\usepackage{a4}
\usepackage{url,epsfig}
\usepackage{lineno}
\newcommand\eps{\varepsilon}

\newcommand\hO{\hat\Omega}
\newcommand\Os{\Omega_s}
\begin{document}
\title{RF gymnastics with transfer matrices}
\author{V. Ziemann\\ 
  Uppsala University, 75120 Uppsala, Sweden}
\date{\today}
\maketitle
\begin{abstract}\noindent
  We introduce transfer matrices to describe the motion of particles in the vicinity
  of the stable and unstable fixed points of longitudinal phase space and use them to
  analyze the transfer of bunches between radio-frequency systems operating at different
  harmonics and voltages. We then apply this formalism to analyze the decoherence
  due to amplitude-dependent tune shift on the asymptotically achievable longitudinal
  emittance and assess tolerances for the injection.
\end{abstract}
%
%
\section{Introduction}
The transfer of bunches between different accelerators requires to carefully adjust
the timing and the energy of the bunches such that their longitudinal emittance is
preserved. Often these accelerators use radio-frequency (RF) systems operating at
different frequencies and voltages, which requires special bunch manipulations
in longitudinal phase space. This is sometimes referred
to as RF gymnastics~\cite{GAROBY} and is usually accomplished by temporarily changing the
amplitude and phase of the RF system in order to rotate the bunch in its longitudinal phase
space. We therefore need to consider the dynamics of particles under the influence of an RF
system operating with peak voltage $\hat V$ and frequency $f_{\mathrm{rf}}=\omega_{\mathrm{rf}}/2\pi.$
Such systems are governed by the equation of motion of the mathematical pendulum
\begin{equation}\label{eq:pendulum}
  \ddot\phi +\Omega_s^2\sin\phi=0
  \qquad\mathrm{with}\qquad
  \Omega_s^2 = -\frac{\omega_{\mathrm{rf}}\eta\cos\phi_s}{\beta^2T_0}\frac{e\hat V}{E_0}
\end{equation}
with the synchrotron frequency $f_s=\Omega_s/2\pi$, the phase slip factor $\eta$, the design
phase $\phi_s$, the speed of the particles $\beta=v/c$, the beam energy $E_0$, and the revolution
time $T_0$. We assume to operate under conditions where $\eta\cos\phi_s<0$ such that the
stable phase is at $\phi=0$. We will always use $\Omega_s$ at the first harmonic and voltage
$\hat V$ as reference and note that the synchrotron tune $\hat\Omega(v,h)$ at voltage $v\hat V$
and harmonic $h$ is related to $\Omega_s$ by $\hat\Omega(v,h)=\sqrt{vh}\Omega_s$. Henceforth we
will always use the lower-case relative voltages in this report.
\par
We assume that during all transitions only one RF system operates at a time, such that
the dynamics longitudinal phase-space is governed by Equation~\ref{eq:pendulum}. A
characteristic feature of the dynamics is the existence of a separatrix, given for the
first harmonic by the equation
\begin{equation}
\dot\phi=\pm 2\Omega_s\cos(\phi/2)\ .
\end{equation}
This curve separates the periodic from the unbounded phase-space trajectories. Moreover,
closed-form solutions of the equations of motion for the pendulum equation, expressed in
terms of Jacobi-elliptic functions~\cite{ABRASTE}, are available~\cite{VZAPB,GITHUB}.
We adapt the MATLAB~\cite{MATLAB} software that accompanies~\cite{VZAPB} to handle different
harmonics $h$ and voltages $v$ and make the software used in this report available
at~\cite{GITHUB}. These analytical solutions are useful in order to explore the limits of
validity of the linear theory that we describe in Section~\ref{sec:lin}. In the following
section we apply the theory to a number of bunch-transfer scenarios. In Section~\ref{sec:deco}
we analyze the decoherence of a mismatched beam, explore its influence on the asymptotically
achievable emittance, and assess tolerances for the injection.
\section{Linear theory}
\label{sec:lin}
If bunches are short compared to the wavelength of the RF systems the
dynamics of the particles is governed by the linearized equation $\ddot\phi +\Omega_s^2\phi=0$
and their motion can be described by the transfer matrix
\begin{equation}\label{eq:TM}
  \hat R(t)=\left(
    \begin{array}{cc} 
      \cos\left(\Omega_st\right) &  \sin\left(\Omega_st\right)/\Omega_s\\
      -\Omega_s\sin\left(\Omega_st\right) & \cos\left(\Omega_st\right)
    \end{array}
  \right)                                      
\end{equation}
that operates on the state of $(\phi,d\phi/dt)$ which we already used in~\cite{VZL}. It is
related to the relative momentum offset $\delta=\Delta p/p$ via
$\dot\phi=\omega_{\mathrm{rf}}\eta\delta$~\cite{VZAPB}. If we
consider motion at another harmonic number $h$ and voltage $v$, we need to adapt the matrix to
\begin{eqnarray}\label{eq:Rs}
  R_s(h,v,t) &=&
               \left(\begin{array}{cc} 1/h & 0 \\ 0 & 1 \end{array} \right)
  \left(
    \begin{array}{cc}
      \cos\left(\hat\Omega t\right) &  \sin\left(\hat\Omega t\right)/\hat\Omega\\
      -\hat\Omega\sin\left(\hat\Omega t\right) & \cos\left(\hat\Omega t\right)
    \end{array}
  \right)                                                      
  \left(\begin{array}{cc} h & 0 \\ 0 & 1 \end{array}\right)\nonumber\\
  &=&
  \left(
    \begin{array}{cc}
      \cos\left(\sqrt{vh}\Omega_st\right) &  \sin\left(\sqrt{vh}\Omega_st\right)/\sqrt{vh^3}\Omega_s\\
      -\sqrt{vh^3}\Omega_s\sin\left(\sqrt{vh}\Omega_st\right) & \cos\left(\sqrt{vh}\Omega_st\right)
    \end{array}
  \right)\ ,
\end{eqnarray}
where we use the abbreviation $\hat\Omega=\sqrt{vh}\Omega_s$. The scaling with $v$ and $h$ is
obvious from the definition of $\Omega_s$ in Equation~\ref{eq:pendulum}. The two outer matrices
in the first equality are necessary to scale the phase at harmonic $h$, because there are $h$
buckets in the longitudinal extent where there is only one bucket at the first harmonic.
Essentially we stretch the phase-axis first, then we rotate with frequency $\hat\Omega$, and
finally we transform back into the phase space of the first harmonic. We point out that the
small-angle synchrotron period $\hat T$ at harmonic $h$ and voltage $v$ also depends on $v$ and
$h$ and is given by $\hat T=2\pi/\hat\Omega=2\pi/\sqrt{vh}\Omega_s$.
\par
Moreover, also in the vicinity of the unstable fixed point the dynamics is almost linear and is
described by $\ddot\psi -\Omega_s^2\psi=0$, which leads to solutions that depend on hyperbolic
rather than trigonometric functions. The corresponding transfer matrix thus becomes
\begin{equation}\label{eq:Ru}
  R_u(h,v,t)=
  \left(
    \begin{array}{cc}
      \cosh\left(\sqrt{vh}\Omega_st\right) &  \sinh\left(\sqrt{vh}\Omega_st\right)/\sqrt{vh^3}\Omega_s\\
      \sqrt{vh^3}\Omega_s\sinh\left(\sqrt{vh}\Omega_st\right) & \cosh\left(\sqrt{vh}\Omega_st\right)
    \end{array}
  \right)\ .
\end{equation}
In the next section we will apply these transfer matrices to analyze bunch transfers and explore
the limits of validity of the approximations.
\section{Applications}
\label{sec:appl}
\subsection{Equilibrium beam}
The beam matrix $\sigma(h,v)$ that is invariant under mapping with $R_s$ describes the matched
beam---the longitudinal equilibrium distribution. We therefore try to find a beam matrix of the
form
\begin{equation}
\sigma(h,v)=\left(\begin{array}{cc} a & 0 \\ 0 & b \end{array}\right)
\end{equation}
and require $\sigma(h,v)=R_s\sigma(h,v)R_s^{\top}$. Evaluating the product of the three matrix
elements and considering the $12$-element, we find
\begin{equation}
  0=\sin(\phi)\cos(\phi)\left[-ah\hat\Omega+\frac{b}{h\hat\Omega}\right]
  \qquad\mathrm{or}\qquad
  \frac{b}{a}=h^2\hat\Omega^2
\end{equation}
where the equality has to be satisfied for all $\phi=\hat\Omega t$ with $\hat\Omega=\sqrt{vh}\Omega_s$.
Equating the other matrix elements leads to the same condition for $b/a$. A suitable choice for
$\sigma(v,h)$ is therefore proportional to
\begin{equation}\label{eq:sig}
\sigma(h,v)=\left(\begin{array}{cc}1/h\hO & 0\\0&h\hO\end{array}\right)
=\left(\begin{array}{cc}1/\sqrt{vh^3}\Omega_s & 0\\0&\sqrt{vh^3}\Omega_s\end{array}\right)\ ,
\end{equation}
where we replaced the abbreviations by $h$, $v$, and $\Omega_s$. The determinant of $\sigma(v,h)$
is unity, such that we have to multiply with the longitudinal emittance in order to obtain a
beam matrix that has a given bunch length.
\subsection{Phase jump to and from the unstable fixed point}
\label{sec:pj}
%
\begin{figure}[tb]
\begin{center}
\includegraphics[width=0.46\textwidth]{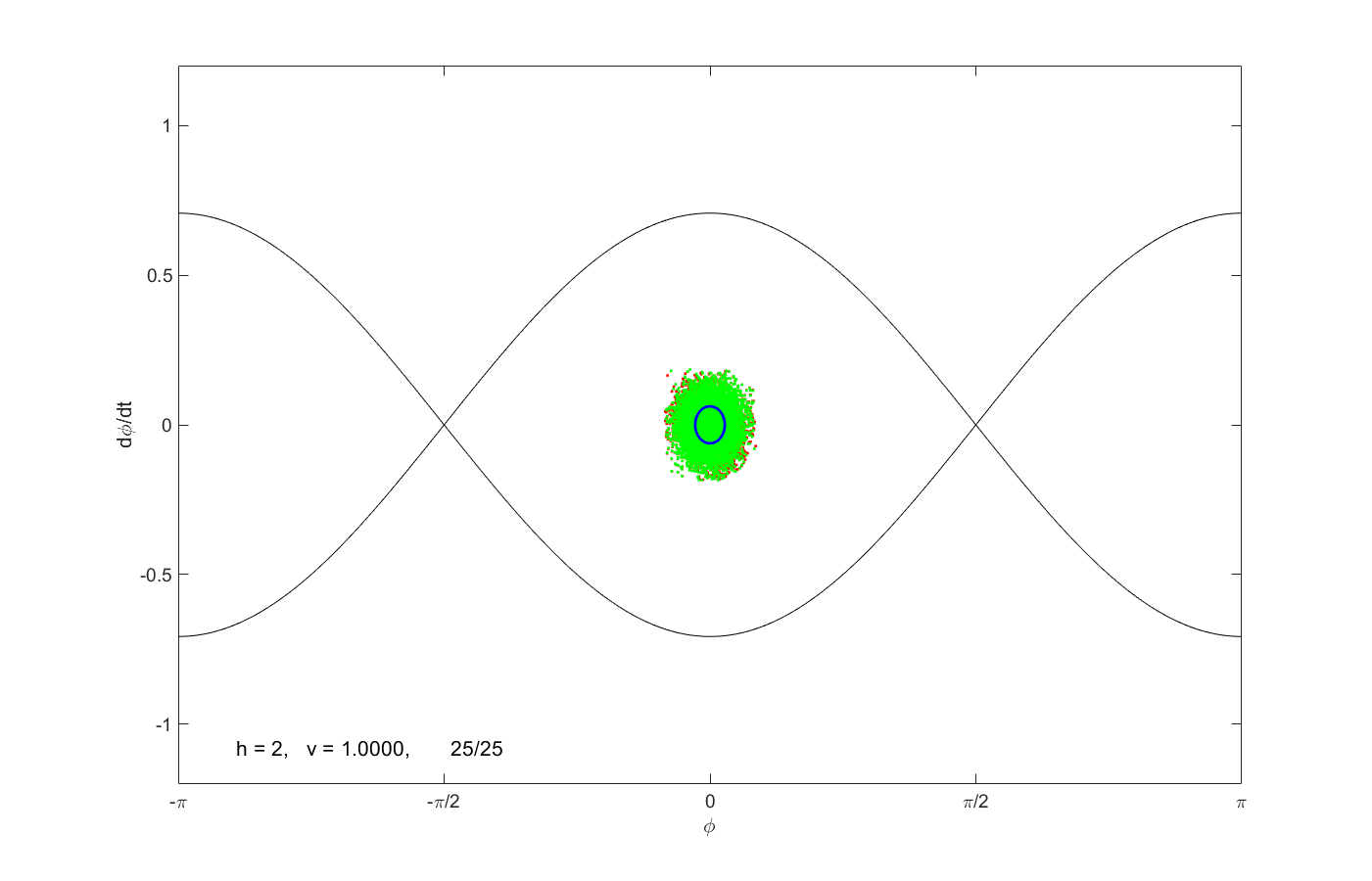}
\includegraphics[width=0.46\textwidth]{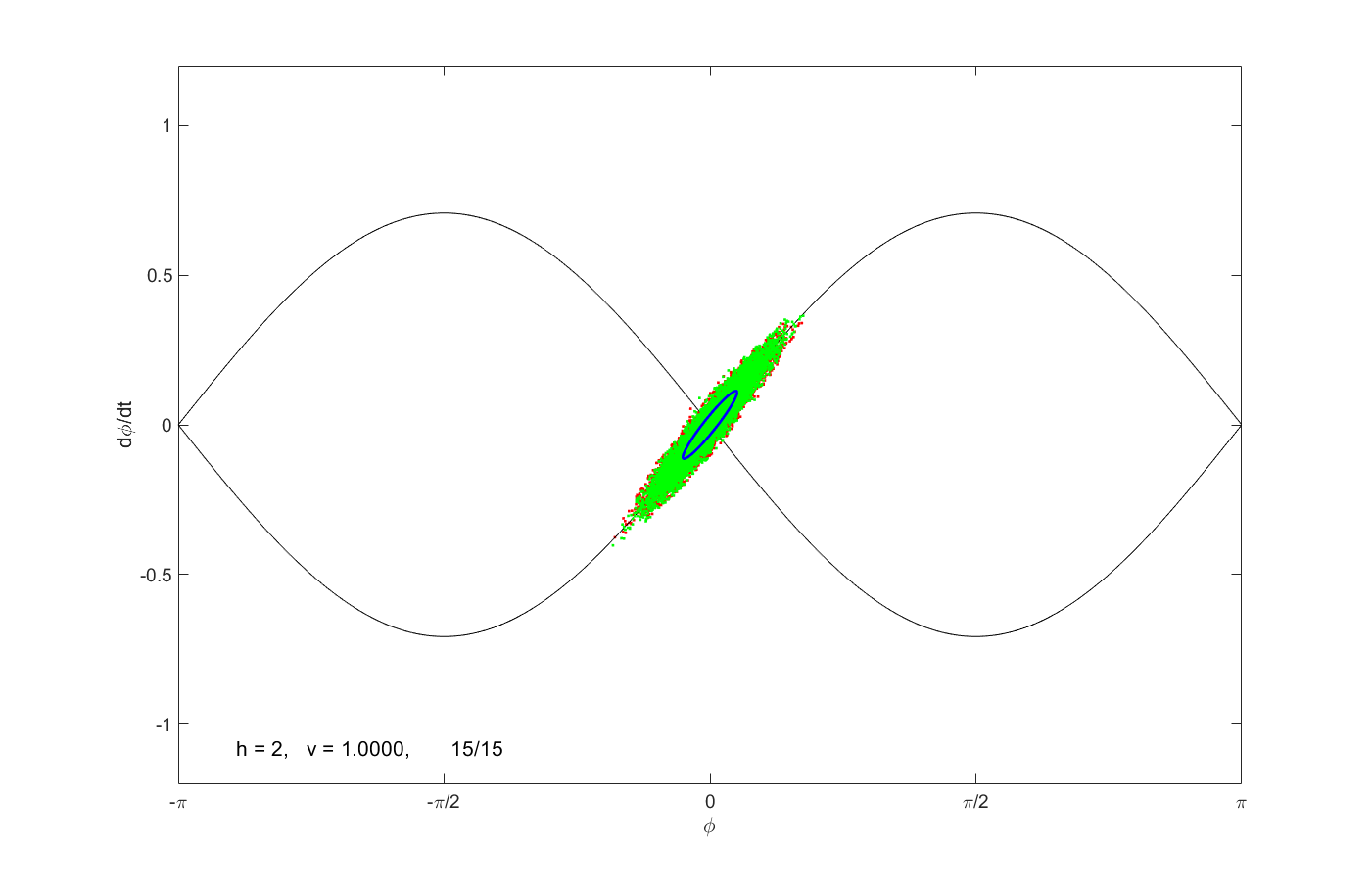}
\includegraphics[width=0.46\textwidth]{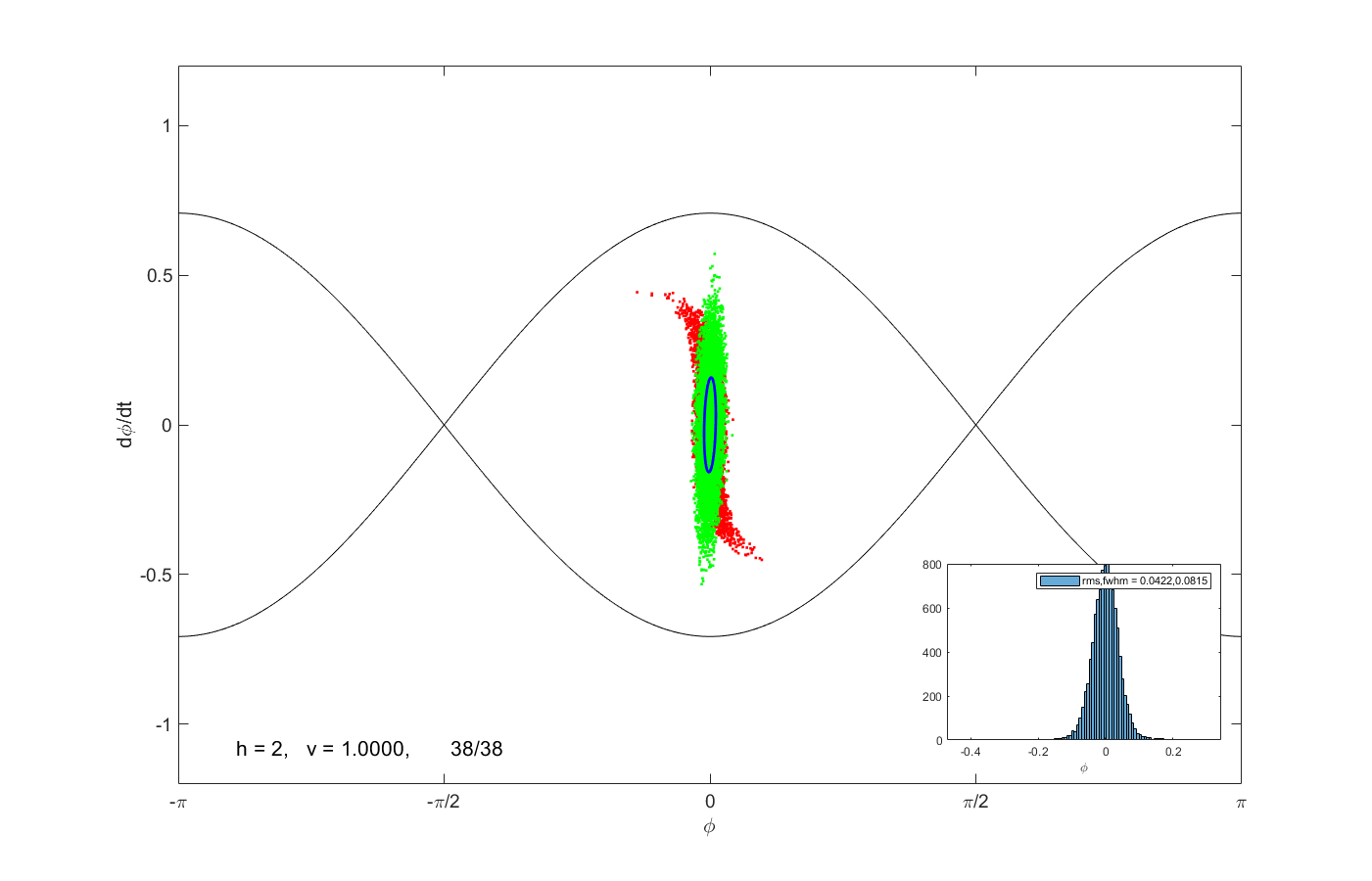}
\includegraphics[width=0.46\textwidth]{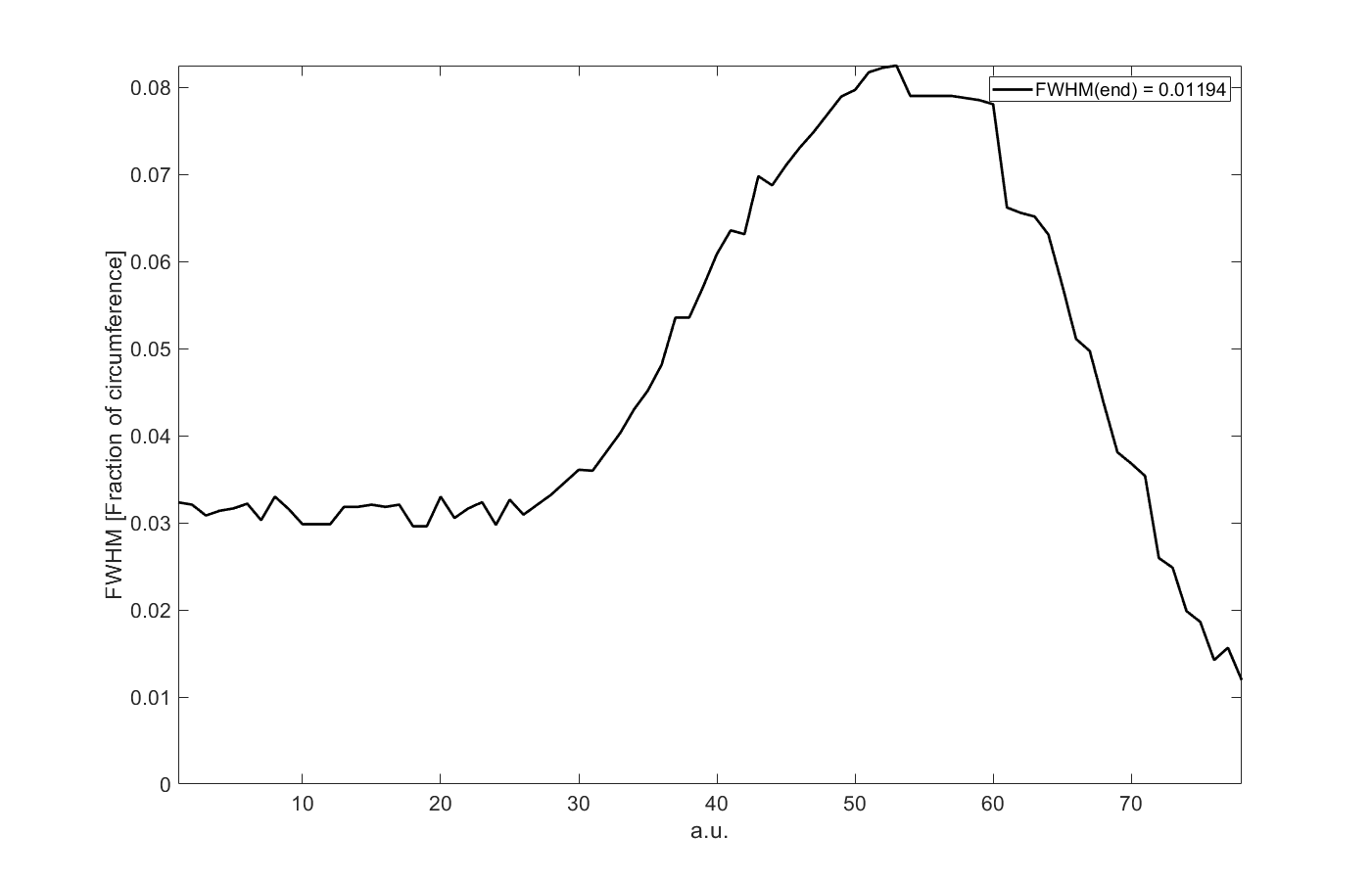}
\end{center}
\caption{\label{fig:sched6}The evolution of the harmonic (top left), the voltage (bottom left)
  and the bunch length (right) for the matched voltage step.}
\end{figure}
In order to explore the validity of the linearized theory, we consider a system in equilibrium,
which is shown on the top left image in Figure~\ref{fig:sched6}. Here the RF system operates on
the second harmonic $h=2$. The red dots show 10000 particles that are propagated with the
full non-linear equations, the green dots show the particles with the same initial conditions,
but propagated with $R_s$. Note that the clouds of red and green dots are very similar. The
blue ellipse is the one-sigma contour of a Gaussian distributions with the rms values used
for the distribution of particles. It appears to be smaller than the cloud of red
and green dots, because the large number of particles saturates the plot. The top-right
image shows the bunch distribution after moving the RF phase by $90^o$ and evolving the
bunch distribution for $t_u=0.15\hat T$ on the unstable fixed point with $R_u$ from
Equation~\ref{eq:Ru}. This leads to an elongation of the bunch along the separatrix. Note
that the red and green dots still agree very well and that the one-sigma contour, which evolves
with $\sigma=R_u\sigma R_u^{\top}$, also describes the distribution quite well. Finally, the
phase of the RF system is restored to the original value and the bunch starts rotating. After
it has rotated for $135^o$ degrees the bunch becomes upright, which is shown on the bottom-left
image. Note that now a discrepancy between the red and green dots appears, because the oscillation
frequency of the large-amplitude particles is lower than that of particles in the center and leads to the
emerging spiral arms, visible underneath the green dots. The blue ellipse also assumes an upright
orientation. The image on the lower right shows the evolution of the full-width at half maximum
(FWHM) of the bunch length, determined from the red dots, with time. We observe that the ratio of
the final to initial bunch lengths is about 0.43. We use the FWHM to quantify the bunch length,
because it is less dependent on the tails of the distribution than the root-mean square.
\par
The good agreement between the distributions of red dots from the full numerical simulation and
the linearized distribution represented by the blue ellipse allows us to use the latter and
derive equations to describe the bunch shortening as a function of the dwell time $t_u$ on the
unstable and $t_s$ on the stable fixed point. The transfer matrix that describes these two
parts is given by
\begin{eqnarray}
  \tilde R&=&
  \left(\begin{array}{cc} 
     \cos(\phi_s) & \sin(\phi_s)/h\hO \\ -h\hO\sin(\phi_s) &  \cos(\phi_s)
  \end{array}\right)
  \left(\begin{array}{cc} 
     \cosh(\phi_u) & \sinh(\phi_u)/h\hO \\ h\hO\sinh(\phi_u) &  \cosh(\phi_s)
  \end{array}\right)\nonumber\\
&=&
  \left(\begin{array}{c}  
    \cos(\phi_s)\cosh(\phi_u)+\sin(\phi_s)\sinh(\phi_u) \\
    -h\hO \sin(\phi_s)\cosh(\phi_u)+h\hO\cos(\phi_s)\sinh(\phi_u)    
 \end{array}\right. \dots \\
&&\qquad\qquad\left.\begin{array}{c}
 \cos(\phi_s)\sinh(\phi_u)/h\hO+\sin(\phi_s)\cosh(\phi_u)/h\hO\\
 \cos(\phi_s)\cosh(\phi_u)-\sin(\phi_s)\sinh(\phi_u)
  \end{array}\right)\nonumber
\end{eqnarray}
where we use the abbreviations $\phi_u=\hO t_u$ and $\phi_s=\hO t_s$. In order to facilitate 
the writing, we introduce $\tilde R_{11}=\cos(\phi_s)\cosh(\phi_u)+\sin(\phi_s)\sinh(\phi_u)$ and
$\tilde R_{12}=\cos(\phi_s)\sinh(\phi_u)/h\hO+\sin(\phi_s)\cosh(\phi_u)/h\hO$. 
\par
If we start with a matched beam distribution, which is proportional to Equation~\ref{eq:sig},
use the transfer matrix $\tilde R$, and calculate the $\tilde\sigma_{11}$ of the matrix
$\tilde\sigma=\tilde  R\sigma(h,v)\tilde R^{\top}$ we find 
\begin{eqnarray}\label{eq:sigmatilde}
\tilde\sigma_{11} &=& \frac{1}{h\hO}\tilde R_{11}^2 +h\hO\tilde R_{12}^2 \\
&=& \frac{1}{h\hO}\left[\cosh(\phi_u)^2 +\sinh(\phi_u)^2+\sin(2\phi_s)\sinh(2\phi_u)\right]\nonumber
\end{eqnarray}
where we use a number of identities among trigonometric and hyperbolic functions. Finding the
minimum bunch length is a matter of calculating the derivative of $\tilde\sigma_{11}$ with 
respect to $\phi_s$, which leads to $\cos(2\phi_s)=0$ or $\phi_s=(2n+1)\pi/4$. The second 
derivative for even $n$ at these values is negative and describes a maximum, but the odd 
values of $n$ describe minima, of which $\phi_s=3\pi/4$ is the first and explains the $135^o$ 
rotation used for turning the bunch upright in Figure~\ref{fig:sched6}.
\par
Inserting $\phi_s=3\pi/4$ in Equation~\ref{eq:sigmatilde} and after some simplifications, 
we arrive at
\begin{equation}
\tilde\sigma_{11} = \frac{1}{h\hO} \left[\cosh(2\phi_u)-\sinh(2\phi_u)\right]=\frac{1}{h\hO}e^{-2\phi_u}\ .
\end{equation}
Since $\tilde\sigma_{11}$ is the square of the bunch length and $1/h\hO$ was the initial bunch 
length, we find that the bunch length-reduction factor is given by $e^{-\phi_u}$. For $t_u=0.15\hat T$,
used in Figure~\ref{fig:sched6}, we find $e^{-2\pi 0.15}\approx 0.39$ which agrees with the observed 
reduction factor reasonably well. Note also that the momentum spread, the height of the ellipse
increases by the corresponding factor. Finally we see that larger dwell times $t_u$ on the unstable 
fixed point will decrease the achievable bunch length, but for values larger than $t_u=0.15\hat T$ 
the spiral-shaped tails of the final distribution increase. This causes the bunch distribution in 
longitudinal phase space to be no longer Gaussian and puts a limit of the useful bunch-length
reduction achievable with phase jumps.
\subsection{Quarter-wave transformer}
\label{sec:quarter}
In the next example, we consider the transfer of a beam that is matched to a first RF system with 
voltage $v_1$ and harmonic $h_1$. This beam is then transferred by a RF system with harmonic $h_2$ 
and voltage $v_2$ to a third one operating at harmonic $h_3$ and voltage $v_3$. The initial
beam matrix is given by Equation~\ref{eq:sig} 
\begin{equation}\label{eq:mb}
\sigma(h_1,v_1)=\left(\begin{array}{cc}1/\sqrt{v_1h_1^3}\Omega_s & 0\\0&\sqrt{v_1h_1^3}\Omega_s\end{array}\right)\ .
\end{equation}
and the final beam has the same shape with index~1 replaced by index~3. Furthermore we chose the
time for the transfer $t_2$ such that the beam performs a quarter of a synchrotron oscillation
with oscillation frequency $\Omega_2=\sqrt{v_2h_2}\Os$ and oscillation period $T_2=2\pi/\Omega_2$.
This makes the time $t_2=T_2/4$ and the transfer matrix becomes
\begin{equation}\label{eq:S}
  R_2(h_2,v_2)=\left(\begin{array}{cc}
    0 & 1/h_2\Omega_2 \\ -h_2\Omega_2 & 0
  \end{array}\right)\ .
\end{equation}
The transfer from the first to the final system is then described by $\sigma(h_3,v_3)=R_2\sigma(h_1,v_1)R_2^{\top}$
\begin{eqnarray}
  \sigma(h_3,v_3)&=& 
  \left(\begin{array}{cc} 0 & \frac{1}{h_2\Omega_2} \\ -h_2\Omega_2 & 0 \end{array}\right)
  \left(\begin{array}{cc}\frac{1}{\sqrt{v_1h_1^3}\Omega_s} & 0\\0&\sqrt{v_1h_1^3}\Omega_s\end{array}\right)  
  \left(\begin{array}{cc} 0 & -h_2\Omega_2 \\ \frac{1}{h_2\Omega_2} & 0 \end{array}\right)
\nonumber\\
   &=&  \left(\begin{array}{cc}\frac{\sqrt{v_1h_1^3}}{v_2h_2^3\Omega_s} & 0\\
         0&\frac{v_2h_2^3}{\sqrt{v_1h_1^3}}\Omega_s\end{array}\right)\ .
\end{eqnarray}
Equating the $11$-matrix elements of $\sigma(h_3,v_3)$ with the last expression in the last equation,
we find
\begin{equation}
  \frac{v_2h_2^3}{\sqrt{v_1h_1^3}}\Omega_s= \sqrt{v_3h_3^3}\Omega_s
\end{equation}
and solving for $v_2h_2^3$ we obtain
\begin{equation}
  v_2 h_2^3 = \sqrt{v_1v_3}\left(h_1h_3\right)^{3/2}  \ .
\end{equation}
we see that using the higher harmonic system at the intermediate stage ($h_2=h_3$) requires
a smaller voltage $v_2$ for the transition. For this configuration we obtain
\begin{equation}\label{eq:stepup}
  v_2=\sqrt{v_1v_3} \left(\frac{h_1}{h_3}\right)^{3/2}
  \qquad\mathrm{if}\quad h_2=h_3
\end{equation}
and if we operate the transition on the first harmonic $h_2=h_1$, we find
\begin{equation}
  v_2=\sqrt{v_1v_3} \left(\frac{h_3}{h_1}\right)^{3/2}
  \qquad\mathrm{if}\quad h_2=h_1\ .
\end{equation}
Finally, if the transition happens between systems operating at the the same harmonic $h_1=h_3$, we
recover the condition given in Equation~11 from~\cite{Boussard}.
\par
Since both the initial and final bunch distributions are matched to their respective
RF systems, the ratio of the bunch lengths $\tilde \sigma_i=\sqrt{\sigma_{11}(h_i,v_i)}$ for
$i=3$ and $i=1$ is given by
\begin{equation}
  \frac{\tilde\sigma_3}{\tilde\sigma_1}=\left(\frac{v_1}{v_3}\right)^{1/4} \left(\frac{h_1}{h_3}\right)^{3/4}\ .
\end{equation}
which is consistent by comparing two matched distributions from Equation~\ref{eq:sig}.
\subsection{Bunch muncher}
Instead of using a single intermediate step to transfer the bunch from one configuration
with voltage $v_1$ and harmonic $h_1$ to one with $v_2$ and $h_2$, we can also use a two-step
transfer with two quarter-wave steps at the original harmonic $h_1$, which is described
in~\cite{BM} and is referred to as a {\em bunch munch.} The first quarter-wave step occurs
at voltage $\hat v<v_1$, whereas the second step occurs at the original voltage $v_1$.
This procedure then provides a matched bunch at $v_2$ and $h_2$. The two transfer matrices
$R(h_1,\hat v)$ and $R(h_1,v_1)$ for the two steps are given by Equation~\ref{eq:S}
and their product is
\begin{equation}\label{eq:T}
  M=R(h_1,v_1)R(h_1,\hat v)
  =\left(\begin{array}{cc} -\sqrt{\hat v/v_1} & 0 \\ 0 & -\sqrt{v_1/\hat v}\end{array}\right)\ .
\end{equation}
We note that the top-left element $M_{11}=-\sqrt{\hat v/v_1}$ describes the demagnification
of the bunch length. Equating $M\sigma(h_1,v_1)M^{\top}$ and the equilibrium bunch distribution
$\sigma(h_2,v_2)$ leads to the condition
\begin{equation}
  \sqrt{v_2h_2^3} \Omega_s = \frac{v_1}{\hat v}\sqrt{v_1h_1^3} \Omega_s
  \qquad\mathrm{or}\qquad
  \frac{\hat v}{v_1} =\sqrt{\frac{v_1}{v_2}} \left(\frac{h_1}{h_2}\right)^{3/2}\ ,
\end{equation}
which gives us the voltage $\hat v$ during the munch that is needed to transform the
matched longitudinal phase space of the first RF system to that of the second system. 
\par
As the product of two off-diagonal matrices from Equation~\ref{eq:S} the matrix $M$ is
diagonal. This is analogous to the assembly of an optical telescope consisting of two
lenses with focal length $f_1$ and $f_2$ with adjacent drift spaces having lengths equal
to the respective focal lengths. The matrix for the telescope has $-f_1/f_2$ and its
inverse on the diagonal. Comparing with Equation~\ref{eq:T} we observe that the
voltages take the role of the focal length $f$. The bunch-muncher thus resembles a
telescope in longitudinal phase space.
\par
The bunch length $\tilde \sigma_i=\sqrt{\sigma_{11}(h_i,v_i)}$ changes from one step to the next
by the magnification $|M_{11}|$ from Equation~\ref{eq:T}, such that we obtain
\begin{equation}\label{eq:demag}
  \frac{\tilde\sigma_2}{\tilde \sigma_1}=\sqrt{\frac{\hat v}{v_1}}
  =\left(\frac{v_1}{v_2}\right)^{1/4} \left(\frac{h_1}{h_2}\right)^{3/4}\ .
\end{equation}
which agrees with the final configuration already found in Section~\ref{sec:quarter}. The
reason is, of course, that both initial and final bunch distributions are matched and the
type of transfer with one or with two steps does not matter as long as the transfer
conserves the longitudinal emittance $\eps$. And this is guaranteed, because the transfer
matrices from Equation~\ref{eq:Rs} and~\ref{eq:Ru} have unit determinant, which follows from
\begin{equation}
  \eps_2^2=\det\sigma_2=\det(R\sigma_1 R^{\top}) = \det(R)\det(\sigma_1)\det(R^{\top})
  = \det(\sigma_1) = \eps_1^2\ .
\end{equation}
\subsection{Four-step harmonic cascade}
%
\begin{figure}[tb]
\begin{center}
\includegraphics[width=0.49\textwidth]{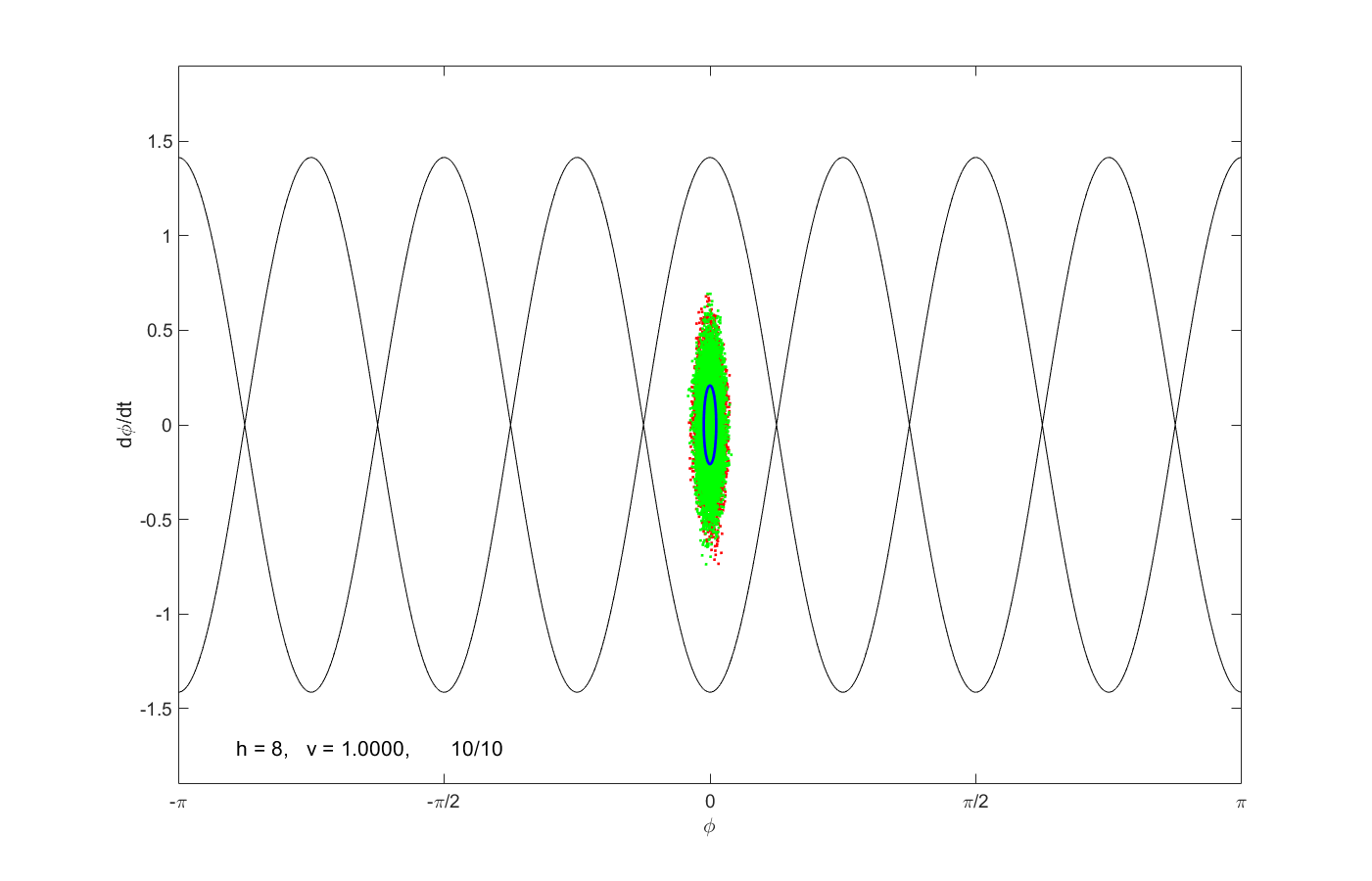}
\includegraphics[width=0.49\textwidth]{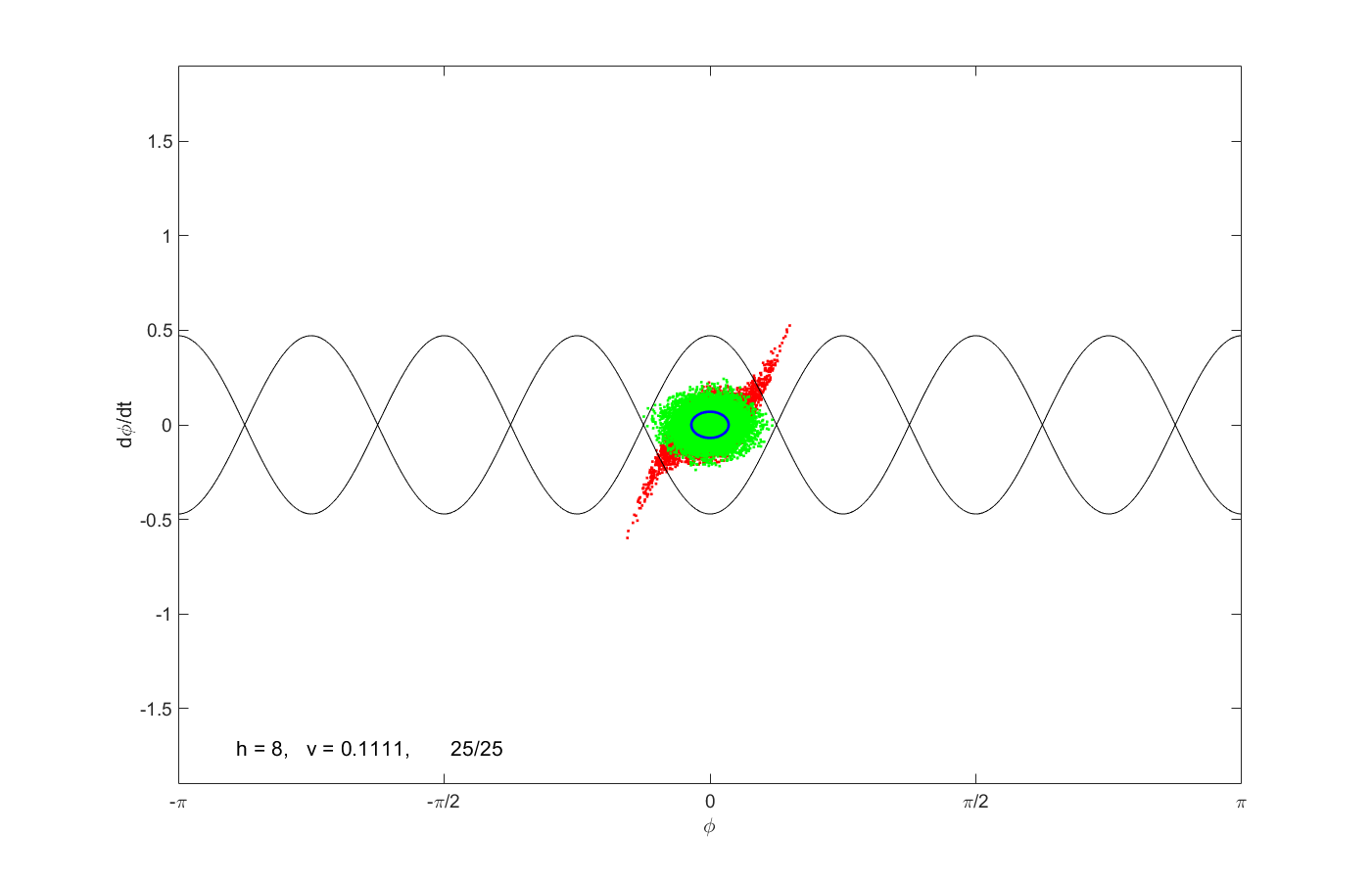}
\includegraphics[width=0.49\textwidth]{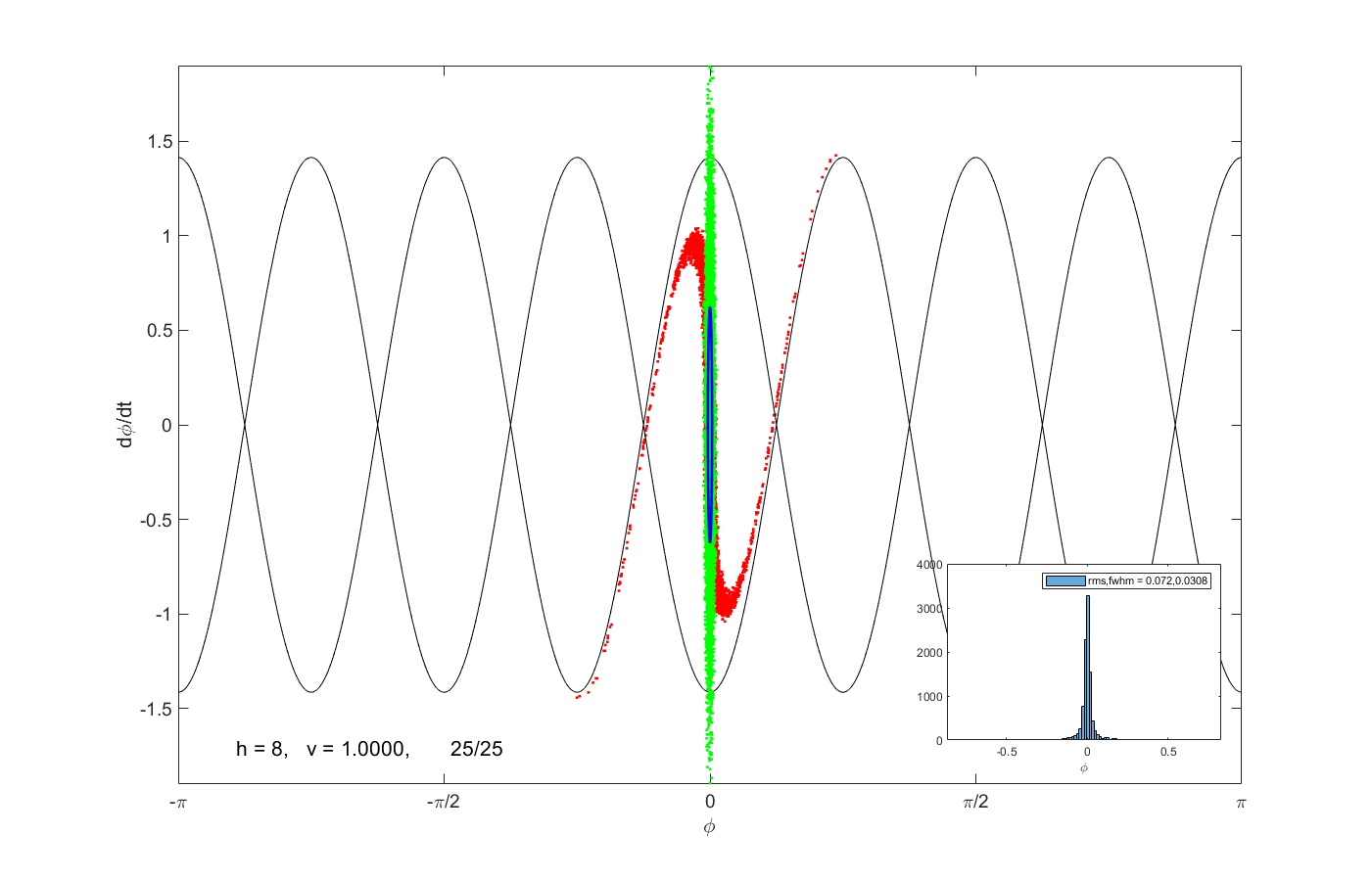}
\includegraphics[width=0.49\textwidth]{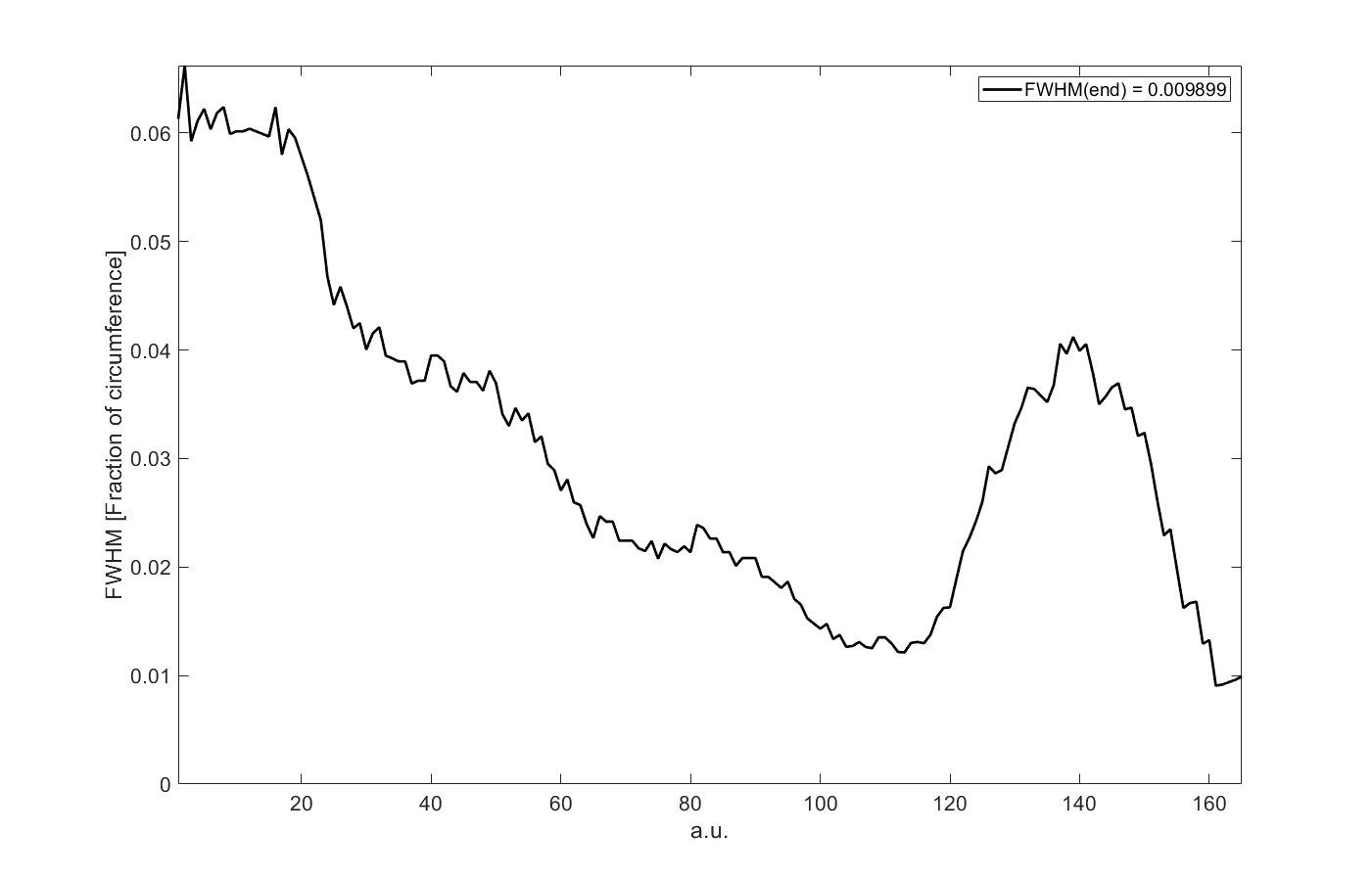}
\end{center}
\caption{\label{fig:sched2}The bunch distribution after the fourth transfer (top left), after
  the the first quarter-wave rotation of the munch munch (top right) and after the the second
  quarter-wave rotation of the munch. The bottom right shows the evolution and the bunch length
  (FWHM).}
\end{figure}
In this example we start from a system with $h_1$ and $v_1$ and then use a quarter-wave
transformer to transfer the bunch to a system with $h_2=2h_1$ operating at the same voltage
$v_2=v_1$. In the next step we double the harmonic again $h_3=4h_1$ while maintaining the
same voltage $v_3=v_1$ and the once again to $h_4=8h_1$ with $v_4=v_1$. As a final step we
apply a munch to reduce the bunch length one last time (albeit without using a
higher-harmonic system). Note that we previously discussed a related cascade, albeit
using two-step transfers based on bunch munches in~\cite{VZL}.
\par
For all quarter-wave steps here, we use the higher-harmonic system operating with the voltage
$\hat v = (1/2)^{3/2}v_1\approx 0.354 v_1$ (from Equation~\ref{eq:stepup}), which ensures
that all transfers are emittance preserving.
The bunch length after the fourth transfer is therefore given by Equation~\ref{eq:demag} and
reads $\tilde \sigma_4=(1/8)^{3/4} \tilde\sigma_1\approx 0.21 \tilde\sigma_1$. For the final munch
we use $\hat v=v_1/9$, such that we hope to obtain a final reduction $M_{11}$ of a factor 3 (from
Equation~\ref{eq:T}), such that the final bunch length should approximately be $\tilde \sigma_f
\approx 0.07\tilde \sigma_1$.
\par
\begin{figure}[tb]
\begin{center}
\includegraphics[width=0.49\textwidth]{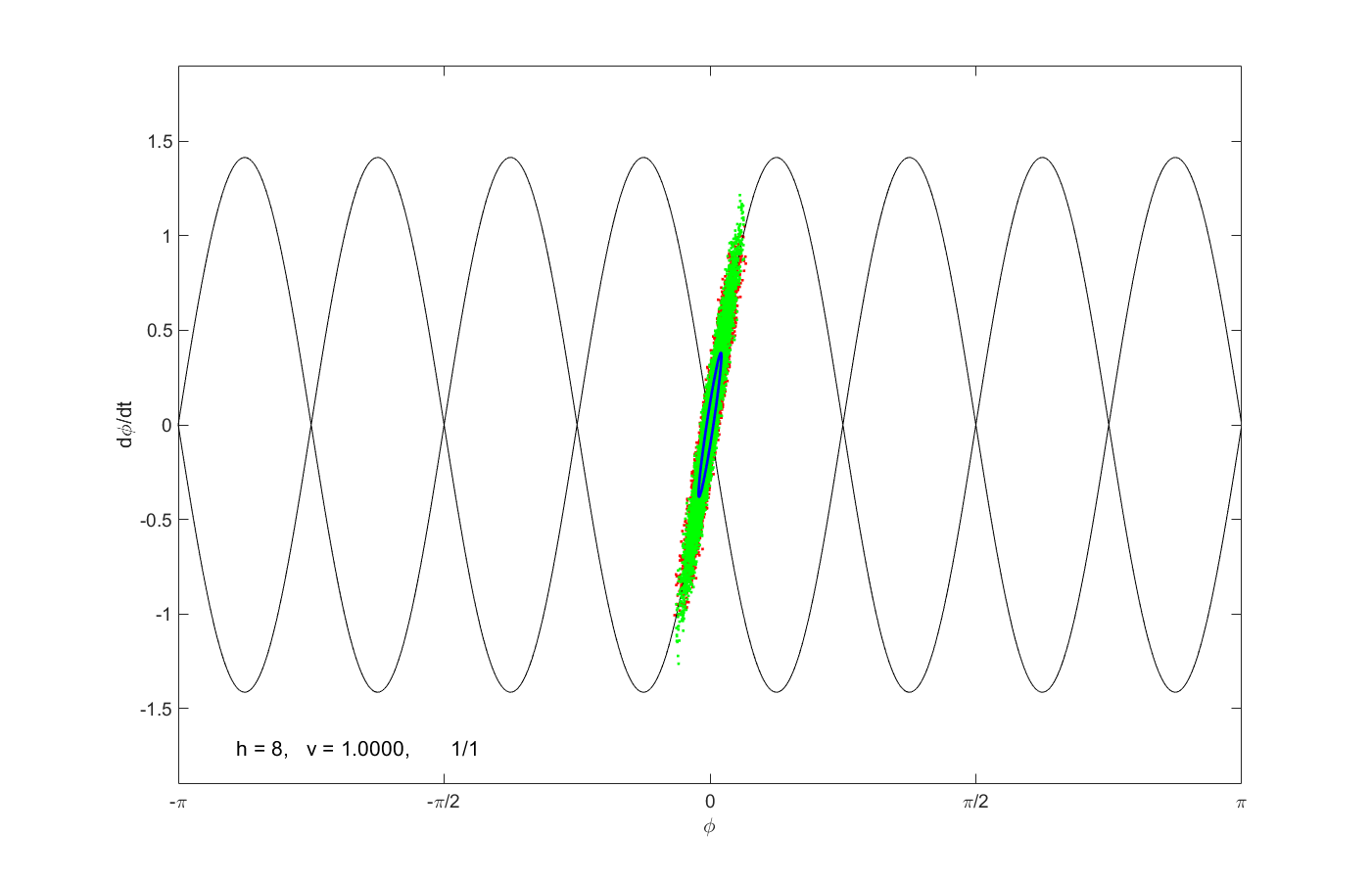}
\includegraphics[width=0.49\textwidth]{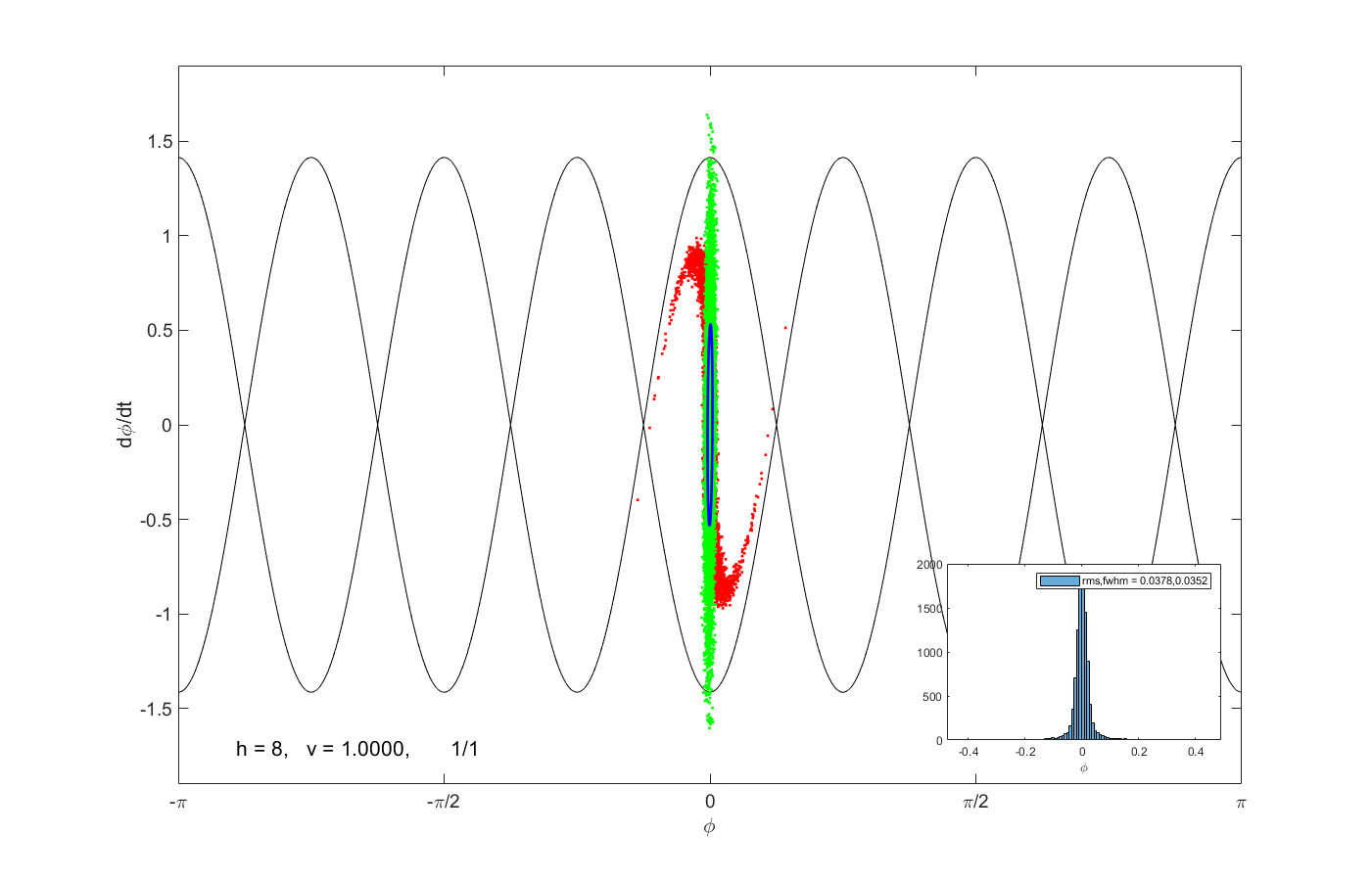}
\end{center}
\caption{\label{fig:withjump}The bunch distribution at the end of the phase jump (left) and
  after the $135^o$ rotation (right).}
\end{figure}
The top left plot in Figure~\ref{fig:sched2} shows the particle distribution at the end of the
four-step cascade. The red dots come from numerically integrating the pendulum equation,
the green dots from the linearized transport, and the blue ellipse from mapping the beam matrix.
We observe that the three descriptions agree very well, considering that the large number of
dots (10000) makes the red and green clouds appear much larger than the rms value of the
distribution that is used to plot the blue ellipse. If we compare the FWHM of the initial
distribution and this one we find that its has decreased by a factor of 5, which is consistent
with the ratio of 0.21, stated in the previous paragraph. On the other hand, the top-right plot,
which depicts  the distribution at the end of the first step of the munch, we observe that the
red dots (numerical) and the green dots (linear transport) differ, because
some of the particles lie close to and even outside of the separatrix. Of course these
particles are outside the domain of validity of the linearized equations. The situation is
aggravated after the second step of the munch, shown on the bottom left in Figure~\ref{fig:sched2}.
The green dots and the blue ellipse show an upright distribution, while the red dots develop
distinct tails, they even reach neighboring phase-space buckets. As a consequence the final
FWHM is about $0.15$ of its initial value rather than $0.07$ times the initial value.
\par
Instead of using a munch to reduce the final bunch length we can also use a phase jump to
stretch the distribution, as shown on the left-hand side in Figure~\ref{fig:withjump}, followed
by a $135^o$ rotation to turn the bunch upright, which is shown on the right-hand image. As in
Section~\ref{sec:pj} we use $t_u=0.15\hat T$, such that we expect the bunch length to shorten
by the factor $e^{-0.15\, 2\pi}\approx0.39$ in this step. The ratio of the FWHM, derived from the
red dots, is about $0.43$,  
which is somewhat larger due to the tails that also develop during the phase jump.
\par
We note that the description of the dynamics with transfer matrices works very well as long as
the bunch distribution stays well inside the stable phase-space area. One thus has to keep
an eye on extreme cases and verify that the distribution stays inside the bucket.
\section{Decoherence and Tolerances}
\label{sec:deco}
%
Up to this point, all bunch transfers between RF systems happen on a time scale of
the synchrotron oscillation periods. In the final RF system, however, they spend a long
time during which the amplitude dependence of the synchrotron frequency causes bunches,
not matched to the RF system, to decohere and increase their emittance. The transfer
matrices introduced earlier in this report allow us to apply the formalism from~\cite{WG}
to calculate this emittance growth. To simplify the notation we specify $h=v=1$ in this
section.
\par
In~\cite{WG} the motion is described in normalized phase space $(x_1,x_2)$ instead of the
physical coordinates, which are $(\phi,\dot\phi)$ in the longitudinal plane. They are
related by
\begin{equation}\label{eq:nps}
  \left(\begin{array}{c} x_1 \\ x_2\end{array}\right)
  = {\cal A}
  \left(\begin{array}{c} \phi \\ \dot\phi\end{array}\right)
  \qquad\mathrm{with}\qquad
  {\cal A} = \left(\begin{array}{cc} \sqrt{\Omega_s} &0 \\ 0 & 1/\sqrt{\Omega_s}\end{array}\right)\ .
\end{equation}
Transforming to normalized coordinates causes the transfer matrix from Equation~\ref{eq:TM}
to become a rotation matrix
and the matched beam matrix $\sigma(h,v)$ from Equation~\ref{eq:sig} to becomes a unit
matrix. Note that the transformation ${\cal A}$ is analogous to the corresponding
transformation in the transverse planes. Exploring this analogy we read off the
longitudinal Twiss parameters $\beta_s=1/\sqrt{\Omega_s}$ and $\alpha_s=0$.
\par
In order to apply the formalism from~\cite{WG}, we determine the amplitude-depen\-dent tune
shift from the synchrotron period $T_p$ for a particle with starting phase $\hat\phi$~\cite{VZAPB}.
It is given by $T_p=(4/\Omega_s)K(\sin\hat\phi/2)$, where
$K$ is a complete elliptic integral~\cite{ABRASTE}. Expanding $K$ to the first order
for small values of $\hat\phi$ gives us the synchrotron frequency
$\Omega_p=2\pi/T_p\approx\Omega_s(1-\hat\phi^2/16)$. We observe that $\Omega_s\hat\phi^2$
is the conserved value of the longitudinal action variable $2J_s=x_1^2+x_2^2$ such that
we find $\Omega_p=\Omega_s-(x_1^2+x_2^2)/16$. In order to make the notation conform
to~\cite{WG}, we multiply with the revolution period $T_0$ in the ring to obtain the
synchrotron phase shift per turn
\begin{equation}\label{eq:phis}
  \phi_s=\Omega_pT_0=\mu_s-\frac{T_0}{16}(x_1^2+x_2^2)
  \qquad\mathrm{with}\qquad
  \mu_s=\Omega_sT_0\ .
\end{equation}
We thus find the coefficient $\kappa=-T_0/16$, which describes the
amplitude dependent tune shift in~\cite{WG}.
\par
The phase space distribution of the incoming beam in normalized phase space is given
by the two-dimensional Gaussian
\begin{equation}\label{eq:gauss}
  \psi\left(\vec x;\vec X,\tilde\sigma\right)=\frac{1}{2\pi\sqrt{\det\tilde\sigma}}
  \exp\left[ -\frac{1}{2}\sum_{j,k=1}^2 \tilde\sigma^{-1}_{jk}(x_j-X_j)(x_k-X_k)\right]\ ,
\end{equation}
where the $X_j$ are the first moments, often called centroids, of the bunch distribution.
In particular, $X_1=\sqrt{\Omega_s}\Delta\phi$ describes the phase offset $\Delta\phi$ of
the injected beam with respect to the zero crossing of the RF in normalized phase space
coordinates. Likewise $X_2=\Delta\dot\phi/\sqrt{\Omega_s}$ is related to the relative
momentum error $\Delta\delta$ through $\Delta\dot\phi=\omega_{\mathrm{rf}}\eta\Delta\delta$ with
the phase-slip factor $\eta$ and the RF frequency $\omega_{\mathrm{rf}}$.
Moreover, $\tilde\sigma={\cal A}\tilde\sigma'{\cal A}^{\top}$, where $\tilde\sigma'$
is given in physical coordinates $\phi$ and $\dot\phi$ 
\begin{equation}
  \tilde\sigma= {\cal A}
  \left(\begin{array}{rr} \sigma_{\phi}^2 & \sigma_{\phi\dot\phi} \\ \sigma_{\phi\dot\phi} & \sigma_{\dot\phi}^2\end{array}\right)
  {\cal A}^{\top}
  =\left(\begin{array}{rr} \Omega_s \sigma_{\phi}^2 &\Omega_s\sigma_{\phi\dot\phi} \\
                 \sigma_{\phi\dot\phi}/\Omega_s &\sigma_{\dot\phi}^2/\Omega_s\end{array}\right)\ .
\end{equation}
Here $\sigma_{\phi}$ is the bunch length and $\sigma_{\dot\phi}=\omega_{\mathrm{rf}}\eta\sigma_{\delta}$ is related
to the relative momentum spread $\sigma_{\delta}$.
\par
As in~\cite{WG} we average the first and second moments of the distribution after
$n$ turns, denoted by a caret. We find the first moments from
\begin{equation}
\hat X_1+i \hat X_2 = \langle e^{-in\phi_s} (x_1+i x_2)\rangle
\end{equation}
with $\phi_s$ given by Equation~\ref{eq:phis} and averaging over the initial distribution
from Equation~\ref{eq:gauss} is denoted by the angle brackets. It turns out that all
integrals can be evaluated in closed form, but we refer to~\cite{WG} for the details of
the  calculation. Likewise, we calculate the second moments after $n$ turns, such as
\begin{equation}
  \langle \hat x_1^2\rangle = \left\langle\left(x_1\cos n\phi_s+x_2\sin n\phi_s\right)^2\right\rangle
\end{equation}
by averaging over the initial distribution from Equation~\ref{eq:gauss}. We point out
that these moments can be calculated for any value of $n$, but here we focus on the
asymptotic emittance that is reached after the initial mismatch has fully decohered
and consider the limit $n\to\infty$. We thus find the asymptotic values of the centroid
$\hat X_1=\hat X_2=0$ and the beam matrix elements $\langle \hat x_1\hat x_2\rangle=0$
and~\cite{WG}
\begin{equation}
  \langle \hat x_1^2\rangle  =  \langle \hat x_2^2\rangle 
  = \frac{1}{2}(\tilde\sigma_{11}+\tilde\sigma_{22})+\frac{1}{2}(X_1^2+X_2^2)\nonumber\\
\end{equation}
where $X_1$, $X_2$, and $\tilde\sigma$ are those appearing in Equation~\ref{eq:gauss}
to define the incoming beam. From these moments it is straightforward to deduce the
longitudinal emittance $\hat\eps$ of the injected beam after it has decohered
\begin{equation}\label{eq:einc}
  \hat\eps=\frac{1}{2}\left[\Omega_s\sigma_{\phi}^2+\sigma_{\dot\phi}^2/\Omega_s\right]+
  \frac{1}{2}\left[\Omega_s\Delta\phi^2+\Delta\dot\phi^2/\Omega_s\right]\ .
\end{equation}
The second term describes the emittance due to injection timing or momentum errors
and the first term is due to the incoming beam matrix. Note that in the absence of timing
and momentum errors an incoming beam with beam matrix proportional to the matched beam
(from Equation~\ref{eq:sig})
\begin{equation}
  \tilde\sigma'
  = \left(\begin{array}{rr} \sigma_{\phi}^2 & \sigma_{\phi\dot\phi} \\
            \sigma_{\phi\dot\phi} & \sigma_{\dot\phi}^2\end{array}\right)
  = \eps_0 \left(\begin{array}{cc} 1/\Omega_s &0 \\ 0 &\Omega_s\end{array}\right)
\end{equation}
leads to an emittance $\hat\eps=\eps_0$. In other words, injecting a matched beam does
not lead to an increased emittance due to decoherence.
\par
%
%
We now use Equation~\ref{eq:einc} to assess tolerances for the injection and model a mismatch of
the incoming beam matrix by assuming an increased bunch length $\sigma_{\phi}=(1+\Delta_a)\sqrt{\eps_0/\Omega_s}$
and correspondingly decreased $\sigma_{\dot\phi}=\sqrt{\eps_0\Omega_s}/(1+\Delta_a)$. The
parameter $\Delta_a$ thus describes the deviation of the aspect ratio of the incoming
beam from the matched configuration. Calculating only the first term in Equation~\ref{eq:einc}
to second order in $\Delta_a$, we obtain the following emittance
$\hat\eps=\eps_0\left(1+2\Delta_a^2\right)$. Moreover, we express the increase of the emittance
due to injection timing and momentum offset in units of the bunch length $\sigma_{\phi}$ and of
$\sigma_{\dot\phi}$ by $\Delta_b=\Delta\phi/\sigma_{\phi}$ and $\Delta_c=\Delta\dot\phi/\sigma_{\dot\phi}$,
respectively. This allows us to write the asymptotic emittance as
\begin{equation}
\hat\eps=\eps_0\left(1+2\Delta_a^2+\frac{\Delta_b^2}{2}+\frac{\Delta_c^2}{2}\right)\ .
\end{equation}
We observe that $\hat\eps$ quadratically increases with the relative deviations from the
optimum values, where an error $\Delta_a$ in the aspect ratio is more significant than
relative phase $(\Delta_b)$ or momentum $(\Delta_c)$ errors.
\section{Conclusions}
In Equation~\ref{eq:Rs} and~\ref{eq:Ru} we introduced transfer matrices to describe the motion
of particles in longitudinal phases space in the vicinity of the stable and the unstable fixed
points. These transfer matrices allow us to treat the longitudinal dynamics with transfer and
beam matrices in much the same way that is commonly used when analyzing transverse dynamics.
We then used this framework to analyze RF gymnastic systems, such as bunch transfer between
different RF systems operating at different harmonics and voltages. The presented linearized
theory makes it possible to derive equations to estimate the performance of RF-gymnastic activities,
like we did in Section~\ref{sec:pj} and we pointed out the limits of validity.
\par
The linearized formalism makes treating amplitude-dependent tune shift feasible and is used
to derive the asymptotic bunch distribution and the resulting emittance due to decoherence.
This is used to assess tolerances for the injection.
\par
Discussions with Heiko Damerau, CERN are gratefully acknowledged.
%
%
\bibliographystyle{plain}

\end{document}